\documentclass[3p,times]{elsarticle}

\long\def\comment#1{ }
\newcommand{\eqn}[1]{Eq.~\eqref{#1}}
\newcommand{\beq}{\begin{equation}}
\newcommand{\eeq}{\end{equation}}
\newcommand{\nn}{\nonumber\\}
\newcommand{\dif}{{\rm d}}
\newcommand{\rmd}{{\rm d}}

\newcommand{\rmi}{{\rm i}}
\newcommand{\rmtr}{{\rm tr}}

\newcommand{\del}{\partial}
\newcommand{\lan}{\langle}
\newcommand{\ran}{\rangle}
\newcommand{\blan}{\big\langle}
\newcommand{\bran}{\big\rangle}

\newcommand{\mcal}{\mathcal}
\newcommand{\wt}{\widetilde}
\newcommand{\bmk}{\bm{k}}
\newcommand{\bmx}{\bm{x}}

\newcommand{\bmu}{\bm{u}}
\newcommand{\bmv}{\bm{v}}
\newcommand{\bmz}{\bm{z}}

\newcommand{\abar}{\bar{\alpha}}

\usepackage{ecrc}

\volume{00}

\firstpage{1}

\journalname{Nuclear Physics A}

\runauth{D.N.~Triantafyllopoulos}

\jid{nupha}

\usepackage{bm,amsmath,amssymb}

\usepackage[figuresright]{rotating}

\begin{document}

\begin{frontmatter}




\title{Multi-gluon correlations in the Color Glass Condensate}


\author{D.N.~Triantafyllopoulos}
\ead{trianta@ectstar.eu}

\address{ECT*, European Centre for Theoretical Studies in Nuclear Physics and Related Areas, Strada delle Tabarelle 286, I-38123 Villazzano (TN), Italy}

\begin{abstract}
The Color Glass Condensate is a universal state of matter which can manifest itself in hadronic processes involving small-$x$ partons, like DIS and $pp$, $pA$ and $AA$ collisions at high energy. Observables are given in terms of multi-gluon correlators, whose ensemble evolves according to a RG equation, the JIMWLK equation. We focus on recent progress towards its solution which lead to quasi-exact, analytic expressions for the multi-gluon correlators at high energy.
\end{abstract}


\end{frontmatter}


In ultra-relativistic hadronic collisions the final state is rather complicated in terms of the type, the number and the distribution of the produced particles. Studying multi-particle correlations is one of the main tools used to extract the dominant physics in the various stages of such a process. For example, significant attention has been recently given to di-hadron correlations in light on heavy hadron collisions, like deuteron-gold at RHIC and the forthcoming proton-lead at the LHC. In the former, when moving towards the deuteron fragmentation region, it is observed an increasing suppression of the azimuthal correlation of the two hadrons when their transverse momenta are a few GeV \cite{Braidot:2011zj,Adare:2011sc}. This is the kinematic regime which encourages the search for parton saturation in the wave-function of the heavy hadron and descriptions of the data based on such a physical mechanism already exist \cite{Albacete:2010pg} or are towards completion \cite{Lappi:2012xe}.

\begin{figure}
 \begin{center}
 \begin{minipage}[b]{0.45\textwidth}
 \begin{center}
 \includegraphics[scale=0.5]{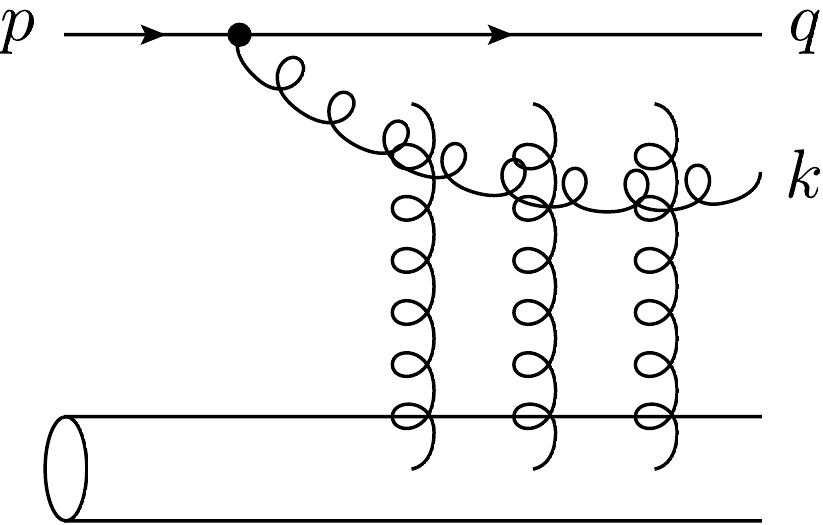}\\(a)
 \end{center}
 \end{minipage}
 \begin{minipage}[b]{0.45\textwidth}
 \begin{center}
 \includegraphics[scale=0.5]{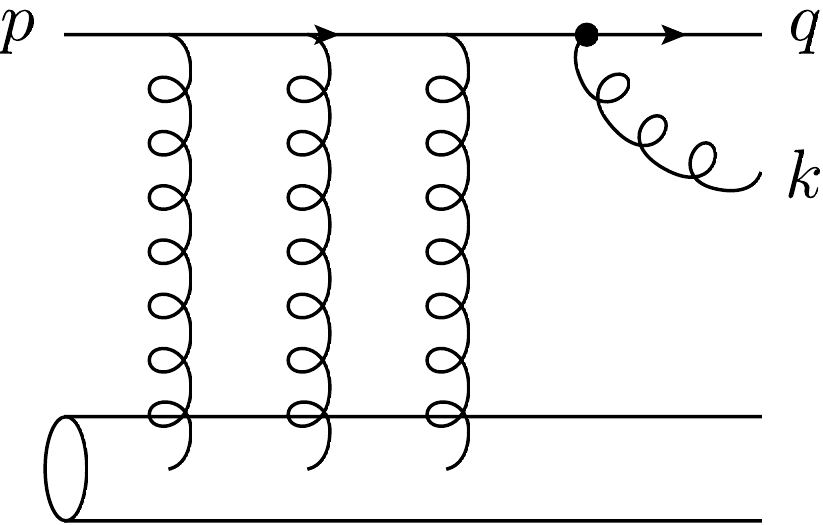}\\(b)
 \end{center}
 \end{minipage}
 \caption{Partonic level diagrams for quark-gluon
 production in high energy $pA$ (or $dA$) collisons 
 and in the proton fragmentation region.\label{fig:dijet}}
 \end{center}
 \end{figure}

Thus we are interested in $hA \to h_1 h_2 X$, with $h$ a dilute projectile hadron, like a proton, and $A$ a dense target. At the partonic level and for example in the case of $qg$ inclusive production, the diagrams are shown in Fig.~\ref{fig:dijet}. In the first a large-$x$ quark from the proton splits into a quark-gluon pair which interacts via multiple gluon exchanges with the small-$x$ part of the target and then is measured in the forward region, while in the second the splitting occurs after the interaction. The target is viewed as a large color field  $\mcal{A}^{\mu}_a = \delta^{\mu+} \alpha_a$, the Color Glass Condensate (CGC) (see e.g.~\cite{Gelis:2010nm}) and the interaction of a parton at transverse position $\bmx$ with such a field is described by Wilson lines like $V^{\dagger}_{\bmx} \equiv {\mbox P}\exp \big[ \rmi g\int \rmd x^-\alpha_a(x^-,\bmx) t^a \big]$. The inclusive $qA \to qgX$ cross section is given by a Fourier transform of \cite{Marquet:2007vb,Mueller:2012bn} 
 \begin{align}\label{cc}
 \frac{N_c^2}{N_c^2-1}
 \bigg\lan \hat{Q}_{\bmx_1\bmx_2\bmx_3\bmx_4} 
 \hat{S}_{\bmx_4\bmx_1}
 -\frac{1}{N_c^2}\,\hat{S}_{\bmx_3\bmx_2}
 -\hat{S}_{\bmu\bmx_1}
 \hat{S}_{\bmx_1\bmx_2}
 + \frac{1}{N_c^2}\, \hat{S}_{\bmu\bmx_2}
 -\hat{S}_{\bmx_3\bmx_4}
 \hat{S}_{\bmx_4\bmv}
 + \frac{1}{N_c^2}\, \hat{S}_{\bmx_3\bmv}
 +\frac{N_c^2-1}{N_c^2}\,\hat{S}_{\bmu\bmv}
 \bigg\ran_Y,
 \end{align}
where $\hat{S}_{\bmx_1\bmx_2} = (1/N_c)\,\rmtr\big({V}^{\dagger}_{\bmx_1} {V}_{\bmx_2}^{\phantom{\dagger}}\big)$ and $\hat{Q}_{\bmx_1\bmx_2\bmx_3\bmx_4} = (1/N_c)\,\rmtr\big({V}^{\dagger}_{\bmx_1} {V}_{\bmx_2}^{\phantom{\dagger}} {V}^{\dagger}_{\bmx_3} {V}_{\bmx_4}^{\phantom{\dagger}}\big)$, with the Wilson lines in the fundamental representation, correspond to the $S$-matrix for the scattering of a dipole and a quadrupole, respectively, off the target color field. It is easy to understand the counting of Wilson lines in \eqn{cc}. For instance, the diagram (a) involves one in the fundamental and one in the adjoint and the latter can be expressed in terms of two fundamental ones. When multiplied with its c.c.~it gives rise to a term with six Wilson lines which is the first term in \eqn{cc} (plus the second term, a $1/N_c^2$ correction). The rapidity $Y$ is determined by the kinematics of the process and in the forward region reads $\exp(-Y) = [|\bmk|\exp(-y_k) + |\bm{q}|\exp(-y_q)]/\sqrt{s}$. The high energy QCD dynamics is contained in the correlators in \eqn{cc}, where the average is taken with the evolved, at rapidity $Y$, target probability distribution $W_{Y}[\alpha]$. This quantum evolution of the GCC obeys the JIMWLK equation (see e.g.~\cite{Iancu:2001ad}), which in its simplest form reads
 \beq\label{JIMWLK}
 \frac{\del W_Y[\alpha]}{\del Y} = 
 -\frac{1}{16 \pi^3} \int_{\bmu\bmv\bmz}
 \mcal{M}_{\bmu\bmv\bmz}
 \left(1 + \wt{V}^{\dagger}_{\bmu} 
 \wt{V}_{\bmv}^{\phantom{\dagger}}
 -\wt{V}^{\dagger}_{\bmu} 
 \wt{V}_{\bmz}^{\phantom{\dagger}}
 -\wt{V}^{\dagger}_{\bmz} 
 \wt{V}_{\bmv}^{\phantom{\dagger}}\right)^{ab}
 \frac{\delta}{\delta \alpha_{\bmu}^a}\,
 \frac{\delta}{\delta \alpha_{\bmv}^b}\,
 W_Y[\alpha] 
 \equiv H W_Y[\alpha] 
 \,\,\,\Rightarrow \,\,\,
 \frac{\del \blan \hat{\mcal{O}}\bran_Y}{\del Y} = 
 \blan H\hat{\mcal{O}}\bran_Y,
 \eeq
where $\mcal{M}_{\bmu\bmv\bmz} = (\bmu-\bmv)^2/[(\bmu-\bmz)^2(\bmz-\bmv)^2]$ is the dipole kernel, the tilde stands for denoting the adjoint representation and the functional derivatives act on the upper and lower end-points of the Wilson lines $V^{\dagger}$ and $V$ respectively. This form holds only when acting on gauge invariant operators and this is the case for those appearing in \eqn{cc}.

\begin{figure}
 \begin{center}
 \begin{minipage}[b]{0.33\textwidth}
 \begin{center}
 \includegraphics[scale=0.47]{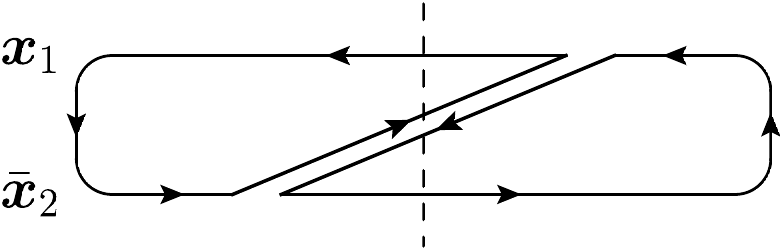}
 \end{center}
 \begin{center}
 \includegraphics[scale=0.47]{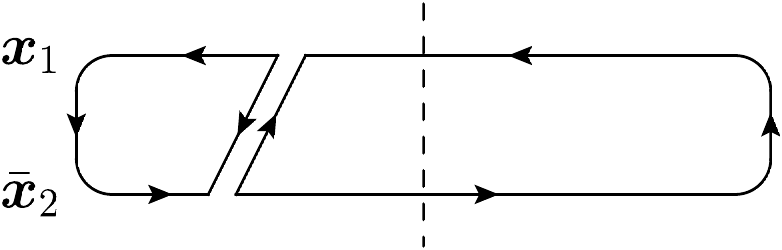}\\(a)
 \end{center}
 \end{minipage}
 \begin{minipage}[b]{0.33\textwidth}
 \begin{center}
 \includegraphics[scale=0.47]{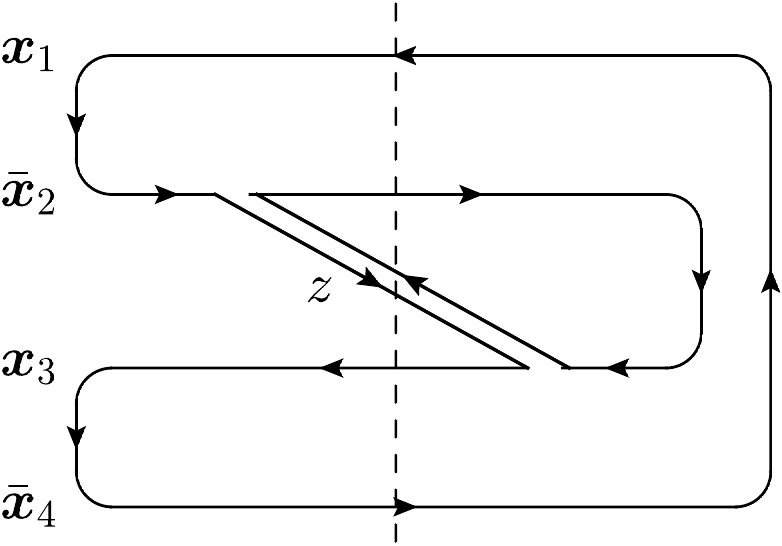} \\(b)
 \end{center}
 \end{minipage}
 \begin{minipage}[b]{0.33\textwidth}
 \begin{center}
 \includegraphics[scale=0.47]{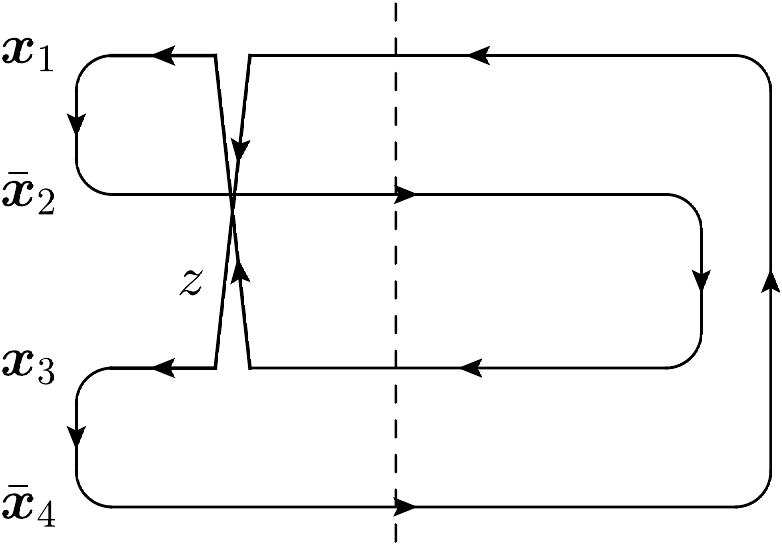}\\(c)
 \end{center}
 \end{minipage}
 \caption{(a) Real and virtual emission of a gluon, a $q\bar{q}$ pair at large-$N_c$, from a dipole. (b) and (c) Two representative diagrams for the corresponding emissions from a quadrupole. In all diagrams the dashed line denotes the interaction with the target.
 \label{fig:dipquad}}
 \end{center}
 \end{figure}

Two popular ways are used in order to calculate the correlators of interest. The one is to reformulate \eqn{JIMWLK} into a Langevin equation \cite{Blaizot:2002xy}, solve directly for the Wilson lines on a lattice and then average over many ``events''. The other is to construct evolution equations for the correlators $\blan \hat{\mathcal{O}} \bran_Y$ of interest according to the second part in \eqn{JIMWLK}. Due to the structure of the real, the last two, terms of the Hamiltonian this method leads to a hierarchy of equations \cite{Balitsky:1995ub}; when $H_{\rm real}$ acts on $n$ Wilson lines, it can lead to $n+2$ of them. This second method had prevailed over the last decade, since one was mostly interested in the 2-point correlator $\blan \hat{S}_{\bmx_1\bmx_2} \bran_Y$ in terms of which we can write, for example, DIS cross sections and single inclusive gluon production in $pA$ collisions. \eqn{JIMWLK} leads to the dipole equation ($\abar = \alpha_s N_c/\pi$)
 \beq\label{BK}
 \frac{\del \blan \hat{S}_{\bmx_1\bmx_2} \bran_Y}{\del Y}=
 \frac{\abar}{2\pi}\, \int_{\bmz}
 \mcal{M}_{\bmx_1\bmx_2\bmz}
 \blan \hat{S}_{\bmx_1\bmz} \hat{S}_{\bmz\bmx_2}
 -\hat{S}_{\bmx_1\bmx_2} \bran_Y,
 \eeq
whose diagrammatic interpretation in terms of an equivalent projectile evolution is shown in Fig.~\ref{fig:dipquad}.a at large-$N_c$. In this limit (and neglecting Pomeron loops \cite{Triantafyllopoulos:2005cn}) one factorizes $\blan \hat{S} \hat{S}\bran_Y = \blan \hat{S} 
\bran_Y \blan \hat{S} \bran_Y$ to get the BK equation \cite{Balitsky:1995ub,Kovchegov:1999yj}, which is closed and its solution is by now well understood semi-analytically and numerically. The saturation momentum $Q_s$ is obtained as the transition line from the region where $\blan \hat{T} \bran_Y = 1- \blan \hat{S} \bran_Y$ is weak to the region where it is close to unity.

However, as evident in \eqn{cc}, less inclusive quantities require the knowledge of higher-point correlators. The quadrupole equation \cite{JalilianMarian:2004da} can also be derived using \eqn{JIMWLK} and looks considerably more complicated than \eqn{BK}
 \begin{align}\label{Qevol}
 \frac{\del \blan\hat{Q}_{\bmx_1\bmx_2\bmx_3\bmx_4} \bran_Y}
 {\del Y} =
 \frac{\abar}{4\pi}  \int_{\bmz} &
 \big[(\mcal{M}_{\bmx_1\bmx_2\bmz} +
 \mcal{M}_{\bmx_1\bmx_4\bmz} -
 \mcal{M}_{\bmx_2\bmx_4\bmz})
 \blan
 \hat{S}_{\bmx_1\bmz}\hat{Q}_{\bmz\bmx_2\bmx_3\bmx_4}
 \bran_Y
 +(\mcal{M}_{\bmx_1\bmx_2\bmz} +
 \mcal{M}_{\bmx_2\bmx_3\bmz} -
 \mcal{M}_{\bmx_1\bmx_3\bmz})
 \blan
 \hat{S}_{\bmz\bmx_2}\hat{Q}_{\bmx_1\bmz\bmx_3\bmx_4}
 \bran_Y
 \nonumber \\*[-0.1cm]
 &\hspace{-3cm}+(\mcal{M}_{\bmx_2\bmx_3\bmz} +
 \mcal{M}_{\bmx_3\bmx_4\bmz} -
 \mcal{M}_{\bmx_2\bmx_4\bmz})
 \blan
 \hat{S}_{\bmx_3\bmz}\hat{Q}_{\bmx_1\bmx_2\bmz\bmx_4}
 \bran_Y
 +(\mcal{M}_{\bmx_1\bmx_4\bmz} +
 \mcal{M}_{\bmx_3\bmx_4\bmz} -
 \mcal{M}_{\bmx_1\bmx_3\bmz})
 \blan
 \hat{S}_{\bmz\bmx_4}\hat{Q}_{\bmx_1\bmx_2\bmx_3\bmz}
 \bran_Y
 \nn
 &\hspace{-3cm}-(\mcal{M}_{\bmx_1\bmx_2\bmz} + 
 \mcal{M}_{\bmx_3\bmx_4\bmz}
 -\mcal{M}_{\bmx_1\bmx_3\bmz} - 
 \mcal{M}_{\bmx_2\bmx_4\bmz})
 \blan
 \hat{S}_{\bmx_1\bmx_2}\hat{S}_{\bmx_3\bmx_4}
 \bran_Y
 -(\mcal{M}_{\bmx_1\bmx_4\bmz} + 
 \mcal{M}_{\bmx_2\bmx_3\bmz}
 -\mcal{M}_{\bmx_1\bmx_3\bmz} - 
 \mcal{M}_{\bmx_2\bmx_4\bmz})
 \blan
 \hat{S}_{\bmx_3\bmx_2}\hat{S}_{\bmx_1\bmx_4}
 \bran_Y
 \nn
 &\hspace{-3cm}-(\mcal{M}_{\bmx_1\bmx_2\bmz} + 
 \mcal{M}_{\bmx_3\bmx_4\bmz}
 +\mcal{M}_{\bmx_1\bmx_4\bmz} + 
 \mcal{M}_{\bmx_2\bmx_3\bmz})
 \blan
 \hat{Q}_{\bmx_1\bmx_2\bmx_3\bmx_4}
 \bran_Y\big].
 \end{align}
All terms have an easy interpretation; for instance the real term $\mcal{M}_{\bmx_2\bmx_3\bmz} \hat{S}_{\bmz\bmx_2}\hat{Q}_{\bmx_1\bmz\bmx_3\bmx_4}$ corresponds to Fig.~\ref{fig:dipquad}.b, while the virtual one $\mcal{M}_{\bmx_1\bmx_3\bmz}\hat{S}_{\bmx_3\bmx_2}\hat{S}_{\bmx_1\bmx_4}$ to Fig.~\ref{fig:dipquad}.c. Even at large-$N_c$, where $\blan \hat{S} \hat{Q}\bran_Y = \blan \hat{S} 
\bran_Y \blan \hat{Q} \bran_Y$ and \eqn{Qevol} becomes a closed inhomogeneous and linear equation (with $\blan \hat{S} \bran_Y$ known from \eqn{BK}), the large number of transverse variables together with the non-locality in the transverse space are prohibitive for the possibility of a  numerical solution. We stress that the quadrupole is a new object and a priori there is no reason to think that it may be written as a product of dipoles.

Thus one takes the Langevin approach to calculate multi-gluon correlators numerically, as done for some configurations in \cite{Dumitru:2011vk}. It was found that the numerical data for the quadrupole are very well-described by a Gaussian mean field approximation (MFA); an extrapolation to arbitrary $Y$ of the McLerran-Venugopalan (MV) model, the typical initial condition at some $Y_0$. Such a Gaussian probability distribution $W_Y[\alpha]$ involves a single kernel, hence all high-point correlators can be expressed in terms of the 2-point one. In fact it was suggested long time ago that a ``random phase approximation'' to JIMWLK leads to a Gaussian Hamiltonian \cite{Iancu:2002aq}, even though no explicit reference was made to high-point functions. A Gaussian ansatz was also used in \cite{Kovchegov:2008mk} to estimate $1/N_c^2$ corrections to the BK equation.

One can prove that a Gaussian approximation is a quasi-exact solution to the JIMWLK equation \cite{Iancu:2011ns,Iancu:2011nj}. At first glance, such a statement looks very peculiar since the JIMWLK Hamiltonian is highly non-linear because of the Wilson lines in \eqn{JIMWLK} and which arise from the propagation of an emitted gluon in the background target field. This non-linearity is indeed present in the dipole equation but not in the quadrupole one which is linear in $\hat{Q}$ (at large-$N_c$). This already suggests that a Gaussian approximation can be a possible solution with all the non-linearities absorbed in its kernel or, equivalently, in the 2-point function. At saturation, where the target is dense, real emissions are suppressed and the virtual part $H_{\rm virt}$ composed of the first two terms of the Hamiltonian dominates. This piece is evidently of Gaussian form, including the second term; those Wilson lines simply transform the ``left'' functional derivatives to ``right'' ones which act on the lower and upper end-points of the Wilson lines $V^{\dagger}$ and $V$ respectively. Now we integrate the dipole kernel over $\bmz$ in the region $1/Q_s \ll |\bmu-\bmz|,|\bmv-\bmz| \ll |\bmu-\bmv|$, with the lower limit imposed by our approximation and the upper one chosen to give the dominant logarithmic contribution $2 \ln\big[(\bmu-\bmv)^2 Q_s^2\big]$. So far this treatment leads to a Hamiltonian valid only at saturation, but performing the same approximation in \eqn{BK} we see that this logarithm is related to the logarithmic derivative of the dipole w.r.t.~$Y$. With such a replacement we arrive at the main result
 \beq\label{HG}
 H_{\rm G} = \frac{1}{4 g^2 C_F}
 \int_{\bmu\bmv}
 \frac{\dif \ln \blan \hat{S}_{\bmu\bmv}\bran_Y}{\dif Y}\,
 \left(\frac{\delta}{\delta \alpha_{{\rm L}\bmu}^a} 
 \frac{\delta}{\delta \alpha_{{\rm L}\bmv}^a}
 +
 \frac{\delta}{\delta \alpha_{{\rm R}\bmu}^a} 
 \frac{\delta}{\delta \alpha_{{\rm R}\bmv}^a}
 \right).
 \eeq
This is a Gaussian Hamiltonian with a kernel which has absorbed the non-linearities and it is most easily determined from the BK equation since in the Gaussian approximation one has $\dif \ln \blan \hat{S}_{\bmu\bmv}\bran_Y/\dif Y = (2 C_F/N_c)\,\dif \ln \blan \hat{S}_{\bmu\bmv}\bran_Y^{\rm BK}/\dif Y$. \eqn{HG} holds at finite-$N_c$, and is correct at saturation by construction and in the  dilute limit as can be inspected.

Using $H_{\rm G}$ one constructs evolution equations for multi-gluon correlators which have the benefit to be local in the transverse plane and are, thus, ordinary differentials equations in $Y$ with $Y$-dependent coefficients. Using a ``separability'' property of the Gaussian kernel in \eqn{HG} one can write the final solution for a correlator as a local function in $Y$. E.g.~the quadrupole at large-$N_c$ with an MV model initial condition (different ones can also be accommodated) reads
 \beq\label{Qsol}
 \blan \hat{Q}_{1234} \bran_Y
 = \frac{\ln\big[\blan \hat{S}_{12}\bran_Y
 \blan \hat{S}_{34}\bran_Y
 /\blan \hat{S}_{13}\bran_Y
 \blan \hat{S}_{24}\bran_Y\big]}
 {\ln\big[\blan\hat{S}_{12}\bran_Y
 \blan \hat{S}_{34}\bran_Y
 /\blan \hat{S}_{14}\bran_Y
 \blan \hat{S}_{23}\bran_Y\big]}\, 
 \blan \hat{S}_{12}\bran_Y
 \blan \hat{S}_{34}\bran_Y
 +
 \frac{\ln\big[\blan \hat{S}_{14}\bran_Y
 \blan \hat{S}_{23}\bran_Y
 /\blan \hat{S}_{13}\bran_Y
 \blan \hat{S}_{24}\bran_Y\big]}
 {\ln\big[\blan\hat{S}_{14}\bran_Y
 \blan \hat{S}_{23}\bran_Y
 /\blan \hat{S}_{12}\bran_Y
 \blan \hat{S}_{34}\bran_Y\big]}\,  
 \blan \hat{S}_{14}\bran_Y
 \blan \hat{S}_{23}\bran_Y
 \eeq
(with $i=\bmx_i$). Formally this is the expression derived in the MV model \cite{JalilianMarian:2004da} and later on generalized at finite-$N_c$ \cite{Dominguez:2011wm}. We notice that the quadrupole above obeys the ``mirror symmetry'' $\blan\hat{Q}_{\bmx_1\bmx_2\bmx_3\bmx_4} \bran_Y = \blan\hat{Q}_{\bmx_1\bmx_4\bmx_3\bmx_2} \bran_Y$, which holds beyond the large-$N_c$ and Gaussian approximations and is a result of symmetry under time-reversal where time stands for $x^{-}$ \cite{Iancu:2011nj}. This symmetry is preserved by JIMWLK due to the two types, left and right, of functional derivatives and implies that the hadron expands symmetrically in the $x^{-}$ direction in the course of evolution. Within the Gaussian approximation only, the quadrupole is also symmetric under the charge conjugation $\blan\hat{Q}_{\bmx_1\bmx_2\bmx_3\bmx_4} \bran_Y = \blan\hat{Q}_{\bmx_2\bmx_3\bmx_4\bmx_1} \bran_Y$.

\begin{figure}
 \begin{center}
 \begin{minipage}[b]{0.45\textwidth}
 \begin{center}
 \includegraphics[scale=0.45]{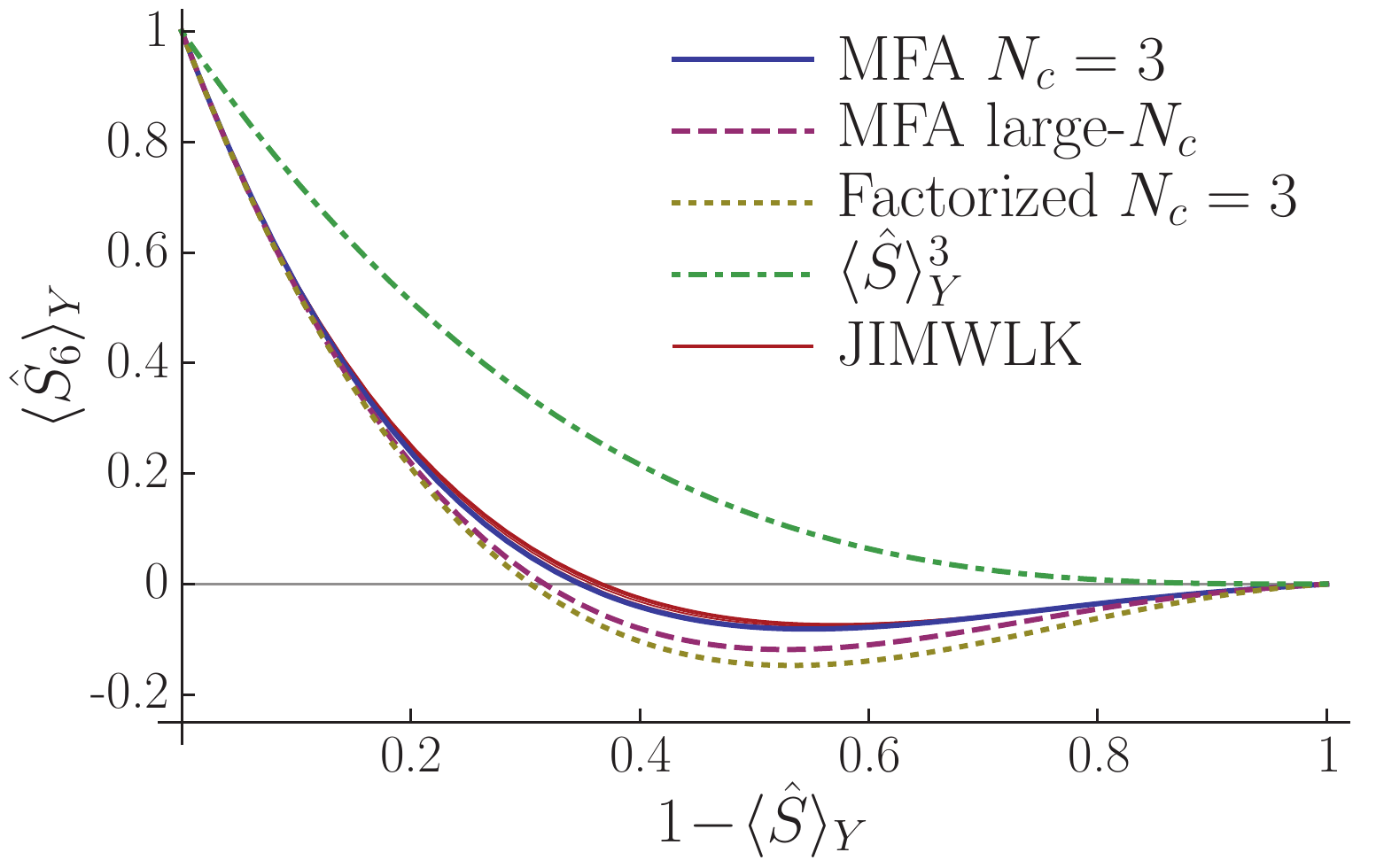}
 \end{center}
 \end{minipage}
 \quad\quad
 \begin{minipage}[b]{0.5\textwidth}
 \begin{center}
 \caption{$\blan \hat{S}_6 \bran_Y$ for the line configuration as a function of $1\!-\!\blan \hat{S} \bran_Y$. Continuous red: JIMWLK with running coupling for six rapidity values from $Y=0$ to $5.18$. Continuous blue: complete MFA (Gaussian approximation) result for $N_c=3$. Dashed magenta: MFA at large-$N_c$. Dotted gold: factorizing the average of $\hat{Q} \hat{S}$ and using MFA at $N_c=3$ for $\lan \hat{Q} \ran_Y$. Dotted dashed green: $\lan \hat{S} \ran_Y^3$, based on just counting Wilson lines. JIMWLK curves from the numerical solution in \cite{Dumitru:2011vk}. MFA curves are analytical expressions in terms of $\lan \hat{S} \ran$, with the latter again provided by \cite{Dumitru:2011vk}. At $Y=0$ JIMWLK and full MFA coincide because of the MV model initial condition. This remains true at any $Y$ in the dilute and in the dense region. A tiny difference, which stabilizes soon, occurs in the transition region.
 \label{fig:S6}}
 \end{center}
 \end{minipage}
 \end{center}
 \end{figure}

At finite-$N_c$ operators mix and one needs to diagonalize a matrix. The quadrupole mixes with two dipoles, while the phenomenologically interesting 6-point operator $\hat{Q}\hat{S}$ in \eqn{cc} mixes with two more operators and the emerging cubic equation has a unique real solution. Simple expressions are obtained for some configurations, like the `line' with $\bmx_3 = \bmx_1$ and $\bmx_4 = \bmx_2$ and thus a single distance determining the value of the correlator, and where we find
 \beq\label{QSfull}
 \blan \hat{Q}\hat{S} \bran_Y =
 \frac{(N_c+2)(N_c-1)}{2 N_c}\,
 \blan \hat{S} \bran_Y^{\textstyle{\frac{3N_c-1}{N_c-1}}}
 -
 \frac{(N_c+1)(N_c-2)}{2 N_c}\,
 \blan \hat{S} \bran_Y^{\textstyle{\frac{3N_c+1}{N_c+1}}}.
 \eeq 
Using the above we can plot $\blan \hat{S}_6 \bran_Y$, defined as the first two terms in \eqn{cc}, and compare with the numerical solution to the JIMWLK equation. As shown in Fig.~\ref{fig:S6} the accuracy for the full curve is excellent at any value of $Y$. Needless to say, analytic expressions for the multi-gluon correlators are of unparalleled significance when, for instance, one wants to performs the Fourier transform of \eqn{cc} and obtain the desired cross section.

\providecommand{\href}[2]{#2}
\begingroup\raggedright

\endgroup

\end{document}